## Clinical Study
# COGEVIS: A New Scale to Evaluate Cognition in Patients with Visual Deficiency

Claire Meyniel [1,2], Dalila Samri,[3] Farah Stefano,[4] Joel Crevoisier,[5] Florence Bonté,[6] Raffaella Migliaccio,[3,7,8] Laure Delaby,[3] Anne Bertrand,[7,9,10] Marie Odile Habert,[11,12] Bruno Dubois,[3,7,8] Bahram Bodaghi,[13] and Stéphane Epelbaum[3,7,10]

[1]AP-HP, Department of Neurophysiology, Hôpital Pitié-Salpêtrière, 75013 Paris, France
[2]Department of Low Vision Rehabilitation, Sainte Marie Hospital, Paris, France
[3]AP-HP, Department of Neurology, Institut de la Mémoire et de la Maladie d'Alzheimer (IM2A), Hôpital de la Pitié-Salpêtrière, 75013 Paris, France
[4]Department of Neurology, Saint-Joseph Hospital, Paris, France
[5]Department of Rehabilitation, MGEN, Maisons-Laffitte, France
[6]Neurology Rehabilitation, Sainte Marie Hospital, USSIF, Paris, France
[7]Sorbonne Universités, UPMC Univ Paris 06, Inserm, CNRS, ICM, 75013 Paris, France
[8]ICM, ICM-INSERM 1127, FrontLab, Paris, France
[9]AP-HP, Department of Neuroradiology, Hôpital de la Pitié-Salpêtrière, 75013 Paris, France
[10]Inria, ARAMIS Project Team, Paris, France
[11]AP-HP, Department of Nuclear Medicine, Hôpital de la Pitié-Salpêtrière, 75013 Paris, France
[12]CATI Multicenter Neuroimaging Platform, Paris, France
[13]AP-HP, Department of Ophthalmology, DHU Vision and Handicaps, Hôpital de la Pitié-Salpêtrière, Paris, France

Correspondence should be addressed to Claire Meyniel; claire.meyniel@aphp.fr

We evaluated the cognitive status of visually impaired patients referred to low vision rehabilitation (LVR) based on a standard cognitive battery and a new evaluation tool, named the COGEVIS, which can be used to assess patients with severe visual deficits. We studied patients aged 60 and above, referred to the LVR Hospital in Paris. Neurological and cognitive evaluations were performed in an expert memory center. Thirty-eight individuals, 17 women and 21 men with a mean age of $70.3 \pm 1.3$ years and a mean visual acuity of $0.12 \pm 0.02$, were recruited over a one-year period. Sixty-three percent of participants had normal cognitive status. Cognitive impairment was diagnosed in 37.5% of participants. The COGEVIS score cutoff point to screen for cognitive impairment was 24 (maximum score of 30) with a sensitivity of 66.7% and a specificity of 95%. Evaluation following 4 months of visual rehabilitation showed an improvement of Instrumental Activities of Daily Living ($p = 0.004$), National Eye Institute Visual Functioning Questionnaire ($p = 0.035$), and Montgomery–Åsberg Depression Rating Scale ($p = 0.037$). This study introduces a new short test to screen for cognitive impairment in visually impaired patients.

## 1. Introduction

Visual impairment, defined as visual acuity of 20/40 or less in the best-corrected better-seeing eye, affects over 280 million people worldwide. Excluding curable etiology such as cataracts or refractive disorders, the most frequent causes are age-related, such as macular degeneration, glaucoma, and diabetic retinopathy [1]. The condition of low vision is therefore strongly age-dependent and affects more than 73% of individuals aged over 65 years [2].

Loss of visual acuity impairs many activities of daily living, including reading, cooking, or selecting clothing/dressing. Associated loss of the peripheral visual field may also cause difficulties for detecting obstacles while walking. Low vision rehabilitation (LVR) delivers multidisciplinary training including visual strategies, occupational therapy, and mobility techniques. Optic aids and nonoptic aids such as tactile marking and signature guides can be used [3]. As the effectiveness of such a multidisciplinary training is difficult to evaluate, few studies have been published on the subject. One study reported a mild improvement of quality of life after LVR [4].

Even though cognitive and visual impairments are both frequent in the elderly, the relationship between the two disorders is still a matter of debate. Several studies have reported an increased cognitive impairment in patients with age-related macular degeneration compared to age-matched subjects [5, 6]. In geriatric health services, the percentage of patients with poor visual acuity was extremely high but patients with visual impairment were found to have lower cognitive scores compared to patients with normal vision [7]. In a large cohort study, the presence of dementia at the time of diagnosis of age-related macular degeneration is not different from what is expected by chance at that age [8]. Moreover, studies assessing the association between dementia and visual impairment are limited due to the fact that many cognitive tests rely on visual skills. Patients presenting a severe loss of vision or blindness can therefore only complete part of the evaluation.

The main objective of this study was the validation of a new scale named the COGEVIS to evaluate the cognitive status of visually impaired patients and therefore detect mild cognitive impairment. The COGEVIS evaluates cognitive function without the use of vision. Our secondary objective was to determine whether LVR was effective in improving quality of life and autonomy among elderly patients with visual impairment.

## 2. Materials and Methods

*2.1. Patients.* We performed a monocentric/single-site prospective study including adults aged 60 and above, referred to the LVR of "Sainte Marie Hospital" of Paris between April 2015 and April 2016. All patients were referred by an ophthalmologist to the LVR outpatient department after diagnosis. Exclusion criteria were (1) previously established diagnosis of neurodegenerative disorder and dementia, (2) ongoing treatment for cancer or other medical illness that would preclude participation, (3) severe psychiatric disorder, and (4) visual acuity of the best eye above 20/70. The study was approved by the French National Research Ethics Committee (CPP, Comité de Protection des Personnes dans la Recherche Biomédicale).

*2.2. Evaluation and Care in the Low Vision Rehabilitation Department.* At the initial patient appointment, a detailed ophthalmologic examination was performed, including visual acuity, monocular manual visual field, and binocular manual visual field. Evaluation of autonomy in the daily living was assessed by a neuropsychologist using Lawton's Instrumental Activities of Daily Living (IADL) [9] The National Eye Institute Visual Functioning Questionnaire (NEI VFQ 25) was used to assess vision-related health status [10]. Depressive symptoms were evaluated with the Montgomery–Åsberg Depression Rating Scale (MADRS) [11]. Cognitive status was evaluated with the COGEVIS (COGnitive Evaluation in VISual impairment), a new scale developed to accommodate impaired vision.

After the first visit, patients followed the low vision rehabilitation program. This consisted of multidisciplinary training performed by optometrists, orthoptists, orientation and mobility instructors, occupational therapists, physiotherapists, and psychologists. Over four months, all patients attended twice-weekly rehabilitation sessions of three hours each. Two hours per week, orthoptists and optometrists worked on improving visual strategies and adjusting devices. Visual strategies were adapted to a patient's functional vision. For example, patients with a central scotoma were trained to use an eccentric fixation, while patients with a constricted visual field were taught how to perform visual scanning. The usefulness of optical devices was also tested: for example, magnifiers were adapted to the smallest readable character size and filters for glare control and lamps were proposed. One hour per week, occupational therapists trained patients to improve their autonomy in activities of daily living. The main domains were cooking, personal care, gesture recognition, self-administered medication, shopping, and financial management. Communication instructors taught patients about computer use for one hour per week. Equipment was adjusted to each patient's vision, such as large print keyboards, magnification software, audio-screen readers, or text-to-speech converters. Orientation and mobility instructors trained patients twice a week to improve walking and mobility autonomy. This work first focused on posture, balance, and foot placement. Depending on the patient's functional vision, training for use of long canes or white canes was proposed, as well as street crossing and public transport autonomy. Psychologists interviewed patients at the beginning of the program and followed patients twice per month throughout the rehabilitation period.

At the end of rehabilitation, the evaluation battery was again carried out, including the IADL, NEI VFQ 25, MADRS, and COGEVIS scales.

*2.3. COGEVIS Description.* COGEVIS (COGnitive Evaluation in VISual impairment) is an assessment measure of cognitive disorders that has the particularity of not soliciting patients' visual abilities. It has been designed to be easily applied in everyday practice by various professionals working with visually impaired patients (e.g., orthoptists and medical doctors).

It is largely composed of subtests derived from global efficiency scales: the Mini Mental State Examination (MMSE) [12], the Frontal Assessment Battery (FAB) [13], and a brief evaluation battery of gestural praxis [14], which were adapted to avoid visual modality.

COGEVIS is composed of

   (i) a subtest of memory (learning and (delayed) recall of 3 words from the MMSE),
   (ii) an evaluation of temporospatial orientation (adapted from MMSE),
   (iii) a short computation test (adapted from MMSE),
   (iv) an evaluation of language:
   (a) denomination based on definitions,
   (b) language comprehension (3 orders of the MMSE),
   (v) an evaluation of the executive functioning:
   (a) letter S fluency in 1 min (from the FAB),
   (b) similarities (from the FAB),
   (vi) an evaluation of the ideomotor apraxia (Mahieux battery) [14],
   (vii) a tactile recognition (naming) test.

COGEVIS is thus a comprehensive cognitive evaluation tool that does not rely on the visual ability to be performed. The scale has a score range of 0–30, with higher scores indicating better function.

The exact French version of COGEVIS (with the verbal instructions and quotation system) and an English translation of it are provided in Supplementary file number 1.

*2.4. Cognitive Status Categorization at the Institute of Memory and Alzheimer's Disease.* Between the initial appointment and the end of the first month of rehabilitation, patients were evaluated at the Institute of Memory and Alzheimer's Disease (IM2A) of the Pitié-Salpêtrière Hospital in Paris. Evaluation included a battery of neuropsychological tests and a consultation with a senior consultant neurologist specializing in cognitive disorder. The neuropsychological battery involved only tests that could be performed by individuals with a visual deficiency, that is, relying more on auditory-verbal skills than on vision. The assessment was composed of the MMSE, the FAB, the digit span forward and backward, lexical (words starting with P) and categorical (animal names) verbal fluencies, the free and cued selective memory test, analysis of praxis, and the California Verbal Learning Test [15]. At the end of the neuropsychological tests and neurological consultation, a consensual diagnosis was made to determine (1) if the participant had normal cognition, mild cognitive impairment (MCI [16]), or a major neurocognitive disorder based on the DSM-V and (2) in the case of cognitive impairment, what was the most probable underlying cause for it. Depending on test results, a MRI and/or positron emission topography (PET) was proposed to help confirm diagnosis.

*2.5. Statistics.* Statistical analysis was performed using the StatView System. Descriptive statistics are presented with mean and standard deviation (SD). A comparison of the participants with normal cognition (on the basis of a consensual clinic-neuropsychological evaluation at IM2A) with those with cognitive impairment (MCI + major cognitive disorder) was performed using Student's *t*-test for continuous variables after visually ensuring Gaussian distribution or chi-squared test for binary or categorical variables. Also, we evaluated the performance of the COGEVIS to diagnose cognitive impairment in the studied population by examining the receiver operating characteristic (ROC) curves of this test. The Wilcoxon signed-rank test was performed to compare the evolution of the COGEVIS, MADRS, IADL, and NEI VFQ 25 pre- and postrehabilitation. The level of significance in all analyses was set at $p < 0.05$.

## 3. Results

Thirty-eight subjects from the LVR of Sainte Marie Hospital of Paris were included in the study. Thirty-two participants completed a neurological evaluation at the IM2A, and 24 completed the follow-up evaluation after LVR. Their mean (±standard error of the mean (SEM)) age was $70.3 \pm 1.3$ years. The cohort included 17 (44.7%) women and 21 (55.3%) men. They had studied for $10.4 \pm 0.8$ years and were predominantly right handed (87%). Their visual and cognitive statuses are described in Figure 1. A detailed description of the population including a comparison of the participants with normal cognition with the cognitively impaired ones (MCI + major cognitive disorder) is provided in Table 1. Five patients assessed presented major cognitive disorders; diagnoses included one person with a mixed pathology origin (vascular + AD), 2 with typical AD, and 2 with typical Lewy body dementia (LBD). Administration of the COGEVIS by a neuropsychologist was feasible and quick (less than 10 minutes for all subjects with a mean administration time of 5 minutes in cognitively unimpaired patients). Interestingly, COGEVIS scores were significantly different between the two groups both at baseline and at follow-up evaluation. In addition, while the cognitively normal participant's COGEVIS scores slightly improved between baseline and follow-up evaluation 4 months later, scores of cognitively impaired participants slightly decreased.

Using the ROC curve method to assess the value of COGEVIS to diagnose cognitive impairment results showed an area under the ROC curve of 0.84. The cutoff point that maximized sensitivity and specificity was 24 with a sensitivity of 66.7% and a specificity of 95% (Figure 2).

Finally, among the 24 subjects who completed a follow-up evaluation after LVR, an improvement of cognition (COGEVIS), functional ability (IADL), quality of life (NEI VFQ 25), and depressive symptoms was observed after 4 months of LVR as displayed in Table 2.

## 4. Discussion

The present study responds to an unmet need for an appropriate diagnostic measure of cognitive impairment in patients with visual deficiency. We developed a new cognitive tool, the COGEVIS, based on the combined expertise of cognitive neurologists, neuropsychologists, and LVR specialists. This new scale is the first comprehensive scale to be validated in a cohort of elderly patients with visual deficiency in whom cognitive impairment is underdiagnosed or wrongly attributed to visual impairment [17]. In this study, COGEVIS was able to identify cognitive impairment with a good diagnostic value (area under the ROC curve of 0.84) and was also used to assess cognitive evolution after LVR. Previous studies screened visually impaired patients with only part of the test omitting items that require image processing. The Leipzig Longitudinal Study of the Aged reported results of part of the MMSE with a maximum total score of 22 instead of 30. Validity of the results was limited, restricted to individuals with very high or very low cognitive performance [18].

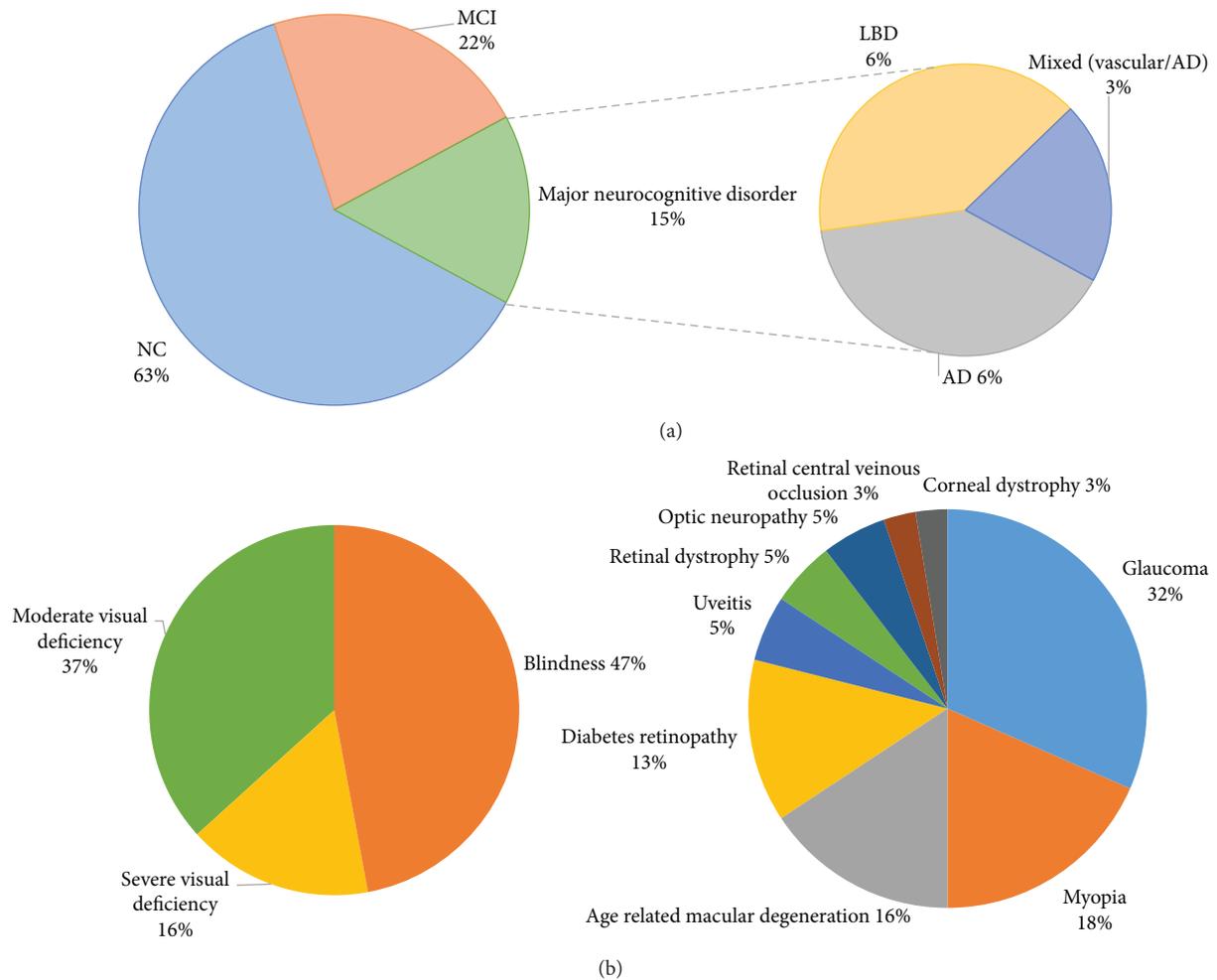

Figure 1: Distribution of the participants according to cognitive status (a) or visual status (b). AD: Alzheimer's disease; LBD: Lewy body dementia; MCI: mild cognitive impairment; NC: normal cognition.

Although the small number of participants did not allow for statistical analysis of major neurocognitive disorder etiology, we qualitatively note that the frequency of cognitive impairment in this population was high, compared to the frequency in the general population [19]. We also found that there was a high percentage of LBD among participants presenting cognitive impairment. Importantly, LBD diagnoses in our study were made according to the latest McKeith et al. criteria [20] and not only proposed in the instance of visual hallucinations. In visual deficiency, there is a high prevalence of visual hallucination related to Charles Bonnet syndrome, a condition in which visual hallucinations develop in association with visual deprivation [21–23]. This condition does not elicit either Parkinsonism or major cognitive fluctuations, two of the three major clinical criteria for LBD. Moreover, we systematically searched for supportive features such as REM sleep behavior disorder, dysautonomia (constipation, orthostatic hypotension), or anosmia to strengthen diagnosis accuracy. However, Charles Bonnet syndrome may not be a benign disease. In Lapid et al.'s study, after an average follow-up time of 33 months, 26% of patients presenting Charles Bonnet syndrome developed dementia. The most commonly diagnosed form of the dementia was LBD [24]. Factors associated with Charles Bonnet syndrome negative outcome were fear-inducing and longer-lasting hallucination episodes associated with a reduction of daily activities [21]. LBD is a disease in which the primary visual cortex is often hypometabolic on fluorodeoxyglucose PET studies [25]. Chronic visual deficiency may induce a vulnerability of the posterior cortex, favoring the development of Lewy

Table 1: Description of the population comparing participants with normal cognition (NC) to cognitively impaired (CI) ones.

| | NC ($N = 20$) | CI ($N = 12$) | $p$ value |
| --- | --- | --- | --- |
| Age | 69.1 (1.7) | 75.0 (2.2) | 0.04 |
| Gender: female | 9 (45) | 7 (58.3) | 0.5 |
| Years of education | 11.8 (1) | 7.9 (1.3) | 0.02 |
| Right handedness | 15 (75) | 12 (100) | 0.12 |
| MMSE | 26.1 (0.8) | 20.1 (1) | 0.0001 |
| FAB | 14.9 (0.6) | 10.8 (0.7) | 0.0001 |
| Digit span forward | 5.8 (0.2) | 4.7 (0.3) | 0.005 |
| Digit span backward | 4.2 (0.2) | 3.2 (0.3) | 0.03 |
| CVLT total recall score | 53.2 (2.4) | 33.2 (2.1) | 0.0001 |
| Intrusions | 0.8 (0.6) | 3.1 (0.7) | 0.02 |
| Recognition | 15.2 (0.5) | 12.6 (0.7) | 0.008 |
| False recognition | 0.6 (0.6) | 4.3 (0.8) | 0.02 |
| Categorical (animals) fluency | 29.0 (2.1) | 18 (2.7) | 0.004 |
| Lexical (letter P) fluency | 21.4 (1.8) | 12.8 (2.4) | 0.008 |
| WAIS-IV vocabulary | 10.7 (0.7) | 7.0 (0.8) | 0.003 |
| Symbolic praxis | 4.9 (0.1) | 4.6 (0.2) | 0.06 |
| Pantomime praxis | 9.8 (0.3) | 8.9 (0.3) | 0.04 |
| MADRS | 13.3 (2.4) | 17.5 (3.3) | 0.3 |
| Baseline COGEVIS | 27.5 (0.6) | 22.9 (0.8) | 0.0001 |
| Follow-up COGEVIS | 28.4 (0.9) | 21.7 (1.1) | 0.0002 |
| Visual deficiency duration (years) | 8.1 (1.9) | 7.6 (2.5) | 0.9 |
| Best-seeing eye visual acuity | 0.13 (0.02) | 0.10 (0.03) | 0.6 |
| IADL | 14.2 (1.2) | 20.6 (1.5) | 0.002 |
| NEI VFQ 25 | 35.8 (2.9) | 28.4 (3.6) | 0.12 |

Values expressed as mean (SEM) and $t$-tests performed for continuous variable or $N$ (%) and chi-squared test performed for categorical variables. CVLT: California Verbal Learning Test; FAB: Frontal Assessment Battery; IADL: Lawton's Instrumental Activities of Daily Living; MADRS: Montgomery–Åsberg Depression Rating Scale; MMSE: Mini Mental State Examination; NEI VFQ 25: National Eye Institute Visual Function Questionnaire; WAIS-IV: Wechsler Adult Intelligence Scale—Fourth Edition.

body dementia. However, this is not substantiated in our study, as neither the degree nor the duration of visual deficiency was associated with cognitive performance. Other factors, which could not be evidenced from this study, could be assessed in a larger cohort of visual deficiency patients followed longitudinally.

The evolution in scores between initial and follow-up assessments after 4 months indicates the efficacy of LVR to improve visually impaired patients' functional abilities and their quality of life. This contributes further evidence in addition to the few studies published on the improvement of quality of life [4] and ongoing utility of LVR in treating patients aged 60 and above. Interestingly, LVR improved cognition according to the significant increase in COGEVIS scores possibly related to the learning of new strategies for planning and organization. We could identify a subgroup of patients with pre-LVR cognitive impairment who did not benefit from LVR, as their post-LVR COGEVIS scores were lower than those at baseline. However, this result does not invalidate LVR in patients with cognitive impairment. Firstly, the low number of cognitively impaired participants does not allow us to draw general conclusions about this result following LVR. Secondly, a specific study, focused only on LVR efficacy in this subgroup of patients, should be conducted in a randomized, placebo-controlled, double blind trial in order to assess the true impact of this therapy. Our study emphasizes that knowing the cognitive status of visually impaired patients before LVR is critical to inform the patient and his family of possible outcomes and adjust expectations regarding LVR.

In previous studies, loss of visual acuity has been reported to be significantly associated with depression [26–28]. Interviews of visually impaired patients older than 60 pointed out the high prevalence of depression in this population (more than 30%) compared to normally sighted peers [29]. In our study, MADRS scores used to assess depressive symptoms showed above average rates of depression among visually deficient patients and higher rates when visual deficiency was associated with cognitive impairment. Depressive symptoms also decreased after LVR. Therefore, discrimination between purely depressive syndromes with cognitive complaints and cognitive impairment due to neurodegenerative diseases is important to adapt the objectives and indication of LVR. COGEVIS could be a suitable test to separate these two syndromes.

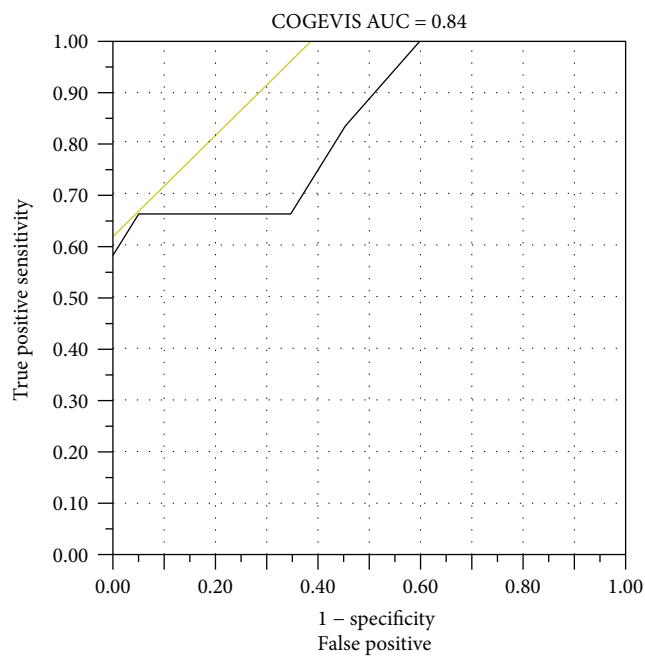

Figure 2: Receiver operating characteristic (ROC) curve of the COGEVIS to diagnose cognitive impairment. AUC: area under the curve. A yellow line is drawn at a 45-degree angle tangent to the ROC curve. This marks a good cutoff point under the assumption that false negatives and false positives have similar costs. In this case, a COGEVIS below 24 yielded a sensitivity of 66.7% and a specificity of 94.7%.

Table 2: Comparison of cognition, functional ability, quality of life, and depression before and after low vision rehabilitation (LVR).

| Mean ± SD | Before LVR | After LVR | Z score | p |
| --- | --- | --- | --- | --- |
| COGEVIS | 24.70 ± 3.98 | 25.65 ± 4.44 | −2.091 | 0.036* |
| IADL | 17.97 ± 6.10 | 15.95 ± 6.56 | −2.896 | 0.004** |
| NEI VFQ 25 | 34.29 ± 15.20 | 38.82 ± 12.74 | −2.092 | 0.035* |
| MADRS | 13.41 ± 9.93 | 9.18 ± 9.67 | −2.090 | 0.037* |

IADL: Lawton's Instrumental Activities of Daily Living; MADRS: Montgomery–Åsberg Depression Rating Scale; NEI VFQ 25: National Eye Institute Visual Function Questionnaire; *$p < 0.05$; **$p < 0.01$.

## 5. Conclusion

COGEVIS is a new, simple, and useful test to screen for cognitive impairment in visually impaired patients. It can also help in the assessment of therapeutic interventions (e.g., LVR) in this population.

## Conflicts of Interest

The authors declare that they have no conflicts of interest.

## Authors' Contributions

Claire Meyniel and Stéphane Epelbaum conceived and drafted the manuscript and are the guarantors for the content. Dalila Samri, Farah Stefano, Joel Crevoisier, Florence Bonté, Raffaella Migliaccio, Laure Delaby, Anne Bertrand, Marie Odile Habert, and Bruno Dubois edited the manuscript.

## Acknowledgments

The authors acknowledge Anna Metcalfe and Sainte Marie Hospital for their help.

## Supplementary Materials

The COGEVIS: a cognitive evaluation tool that does not rely on visual ability. The scale has a score range of 0–30 (higher scores indicating better function). *(Supplementary Materials)*

**Registration: / 3**
"I will read you three words. Repeat them and try to hold them back because I will ask for them again"
1. Cigar / 1
2. Flower / 1
3. Door / 1

**Temporo-spatial orientation: / 5**
"I will ask you some questions about the date of the day and the place where we are."
1. What is the year?  / 1
2. What is the month?  / 1
3. What day of the week?  / 1
4. What is the name of the hospital / clinic / facility where we are?  / 1
5. What city are we in?  / 1

**Attention and calculation: / 3**
"Count backwards from 100 by removing 7 each time until I stop you. 100-7 ... "
1. 93 / 1
2. 86 / 1
3. 79 / 1

**Language: / 8**
"I will ask you a few questions, try to answer as precisely as possible."
1. What object can give the time? (Accepted: watch, clock, pendulum) / 1
2. What object is used when it rains? (Accepted: umbrella, boots, raincoat) / 1

"Listen well and do what I am going to tell you"
3. Take the sheet of paper placed on the table in front of you with your right hand / 1
4. Fold it in half / 1
5. And throw it on the ground / 1

Lexical fluency: letter S (in 1 min):
"Name a maximum of different words starting with the letter S, for example animals, plants ... but no proper names (first names, names of cities or countries)".
Quotation details:  3: ≥ 10 words/   2: 6 to 9 words/   1: 3 to 5 words/   0: 2 words or less    / 3

**Praxis: / 3**
"How do you do with the hand to do ...".
1. Military Salute / 1
2. Send a kiss / 1
3. Drinking a glass / 1

**Touch Recognition: / 2**
"Now you're going to have to recognize objects that I'll put in your hand one by one. What is the name of this object? "
1. Coin / 1
2. Pen / 1

**Recall: / 3**
"Can you recall the words you had to remember earlier? ".
1. Cigar / 1
2. Flower / 1
3. Door / 1

**Executive functions: / 3**
Similarities: "how are they alike ..."
1. A banana and an orange?  / 1
2. A table and a chair?  / 1
3. A tulip, a rose and a daisy?  / 1